\documentclass[journal=jpcafh,manuscript=article,layout=twocolumn,email=false]{achemso}

\usepackage{chemformula} % Formula subscripts using \ch{}
\usepackage[T1]{fontenc} % Use modern font encodings
\usepackage{hyperref}
\usepackage[font=footnotesize]{caption} 
\usepackage{etoolbox}

\apptocmd{\thebibliography}{\footnotesize}{}{}

\SectionNumbersOn

\author{Anton Morgunov}
\author{Henry K. Tran}
\author{Oinam Romesh Meitei}
\author{Yu-Che Chien}
\author{Troy Van Voorhis}
\email{tvan@mit.edu}
\affiliation[MIT]{Department of Chemistry, Massachusetts Institute of Technology, Cambridge, MA}

\title[CEBEs]{\Large MP2-based composite extrapolation schemes can predict core-ionization energies for first-row elements with coupled-cluster level accuracy}

\abbreviations{XPS, XAS, XANES, NEXAFS, RIXS, XES, CBS, CEBE, MP2, CC, DFT, TDDFT, EOM-CC, ADC, X2C, HF, MOM, DIIS, RHF, SCF, RI-MP2, CBS, MAE, RMSE}

\let\oldmaketitle\maketitle
\let\maketitle\relax

\begin{document}

% \begin{tocentry}

% \includegraphics[width=\linewidth]{figures/toc.pdf}

% \end{tocentry}

\twocolumn[
\begin{@twocolumnfalse}
\oldmaketitle
\begin{abstract}
  X-ray photoelectron spectroscopy (XPS) measures core-electron binding energies (CEBEs) to reveal element-specific insights into chemical environment and bonding. Accurate theoretical CEBE prediction aids XPS interpretation but requires proper modeling of orbital relaxation and electron correlation upon core-ionization. This work systematically investigates basis set selection for extrapolation to the complete basis set (CBS) limit of CEBEs from $\Delta$MP2 and $\Delta$CC energies across 94 K-edges in diverse organic molecules. We demonstrate that an alternative composite scheme using $\Delta$MP2 in a large basis corrected by $\Delta$CC-$\Delta$MP2 difference in a small basis can quantitatively recover optimally extrapolated $\Delta$CC CEBEs within 0.02 eV. Unlike $\Delta$CC, MP2 calculations do not suffer from convergence issues and are computationally cheaper, and, thus, the composite $\Delta$MP2/$\Delta$CC scheme balances accuracy and cost, overcoming limitations of solely using either method. We conclude by providing a comprehensive analysis of the choice of small and large basis sets for the composite schemes and provide practical recommendations for highly accurate (within 0.10-0.15 eV MAE) ab initio prediction of XPS spectra.
\end{abstract}
\end{@twocolumnfalse}
]

\section{Introduction}

Developments in the generation of ultra-short X-ray pulses\cite{xraylasers,physicsofxraylasers,xrayroadmap} have sparked a surge of interest in the X-ray Absorption Spectroscopy (XAS), in which details of atomistic structure are revealed by the ionization or excitation of core electrons upon x-ray radiation.\cite{tdxasreview} By measuring the kinetic energy of ionized electrons in an X-ray Photoelectron Spectroscopy (XPS), one can determine the core-electron binding energy (CEBE). Remarkably, these CEBEs are not only element-specific but also are affected by the ionized atom's chemical (and hence electronic) environment, earning XPS the alternative name of electron spectroscopy for chemical analysis (ESCA).\cite{xrsreview} Such specificity enables XPS to perform surface composition analysis,\cite{xpsperspectives,oswald_2013,matthew_2004,watts_wolstenholme_2019} analyze chemical distribution in quantum dots,\cite{zorn_2011} and infer oxidation states and coordination numbers of active centers in catalysis.\cite{nguyen_2019}

The energies at which core electrons are ionized are called edges, with a preceding K, L, or M specifying the principal quantum number corresponding to those core orbitals. Excitations of electrons from core to unoccupied orbitals tend to have similar energies and can be studied through X-ray Absorption Near-Edge Structure (XANES) or Near-Edge X-ray Absorption Fine Structure (NEXAFS). NEXAFS focuses on the very near-edge region (within 10 eV) and reveals the structure and orientation of molecules adsorbed on surfaces.\cite{nexafsreview} While NEXAFS has been traditionally more associated with surface chemistry, it has also been recently applied to study ultrafast chemical dynamics.\cite{nexafsdynamics,electrocyclicnexafs} In contrast, XANES studies regions farther from the edge (within 10-50 eV) and allows to infer oxidation states, coordination number, and types of ligands surrounding the absorbing atom.\cite{xanesreview} XANES has beeen also applied to chemical dynamics on attosecond timescales\cite{dynamicsreview2} and photoinduced proton-coupled electron transfer.\cite{batistapcet}

X-ray spectroscopy is not limited to XAS, however. An electron in an occupied orbital of a core-ionized or core-excited particle may transition into the vacant core orbital, emitting X-ray radiation. Energies of these emitted photons can be studied through X-ray Emission Spectroscopy (XES), which has been recently used to enhance understanding of bonding in transition metals.\cite{xasxeslft} A related Resonant Inelastic X-ray Scattering (RIXS) technique measures the energy and momentum of photons emitted after the core-ionized or core-excited particle undergoes other low-energy transitions (because the emitted photon has a different energy, it has been effectively \textit{scattered}).\cite{rixsreview} Such scattering can provide novel insight into mechanisms of photochemical reactions.\cite{rixstransition}

This paper focuses on the ab initio prediction of CEBE, which is often required for a reliable interpretation of XPS experiments.\cite{xrsreview} Computational methods are benchmarked against experimental CEBEs (also known as core-ionization potentials) aiming to predict experimental values within 0.2 eV.\cite{zheng_cheng_2019} Notably, the experimental resolution of XPS instruments determines the threshold above,\cite{greczynski} implying that even greater accuracy will be expected from theoretical methods as more precise sources of X-ray radiation are developed.

Ejection of a core electron exposes surrounding atoms and electrons to a positive charge, which results in a significant redistribution of electronic density resulting in an increased electronic density in the vicinity of an ionized atom. In order to accurately describe the energetics of this process, a theoretical model must be able to model the changes in molecular orbitals (usually referred to as the relaxation effect) and the effects on electron correlation that this relaxation imposes. Inclusion of relativistic effects and spin-orbit coupling allows to improve the accuracy of predictions especially for heavier elements.\cite{spinorbitxps} There are two conceptually different approaches to calculating core-electron binding energy\cite{xrsreview}. In the first family of methods, response operators modeling the effects of ionization or excitation are applied to ground-state wave functions. These methods allowed successful prediction of energies and physical properties of valence-ionized states,\cite{dreuw_head-gordon_2005} without explicit calculation of the wave function of said ionized states, avoiding the issue of variational collapse to the ground state. Unsurprisingly, significant effort has been made to investigate the applications of response methods to the description of core-ionized states. Time-dependent density functional theory (TDDFT\cite{marques_gross_2004}), a widely used representative of response methods, with commonly used functionals systematically underestimates CEBEs by 10 or more eV, an error typically attributed to the self-interaction and delocalization error characteristic of DFT methods.\cite{hait_2020,hait_head-gordon_2018,yang_2006} These errors can be brought down to 1-3 eV if exchange and correlation functionals are optimized explicitly for core excitations;\cite{lestrange_2015} such calibrations, however, are heavily empiric. Similar accuracy without experimental calibration can be achieved with equation-of-motion coupled-cluster (EOM-CC) methods\cite{eomccsdintro} within a core-valence separation (CVS) approximation.\cite{cederbaum_1980} The inclusion of CVS approximation allows one to skip valence-to-virtual transitions and directly calculate CEBEs, significantly reducing the computational cost of EOM-CC calculations. CVS-EOM-CCSD can describe CEBEs with a mean absolute error (MAE) around 1.75 eV, and CVS-EOM-CCSDT in a quadruple zeta basis set brings down the MAE to 0.15 eV\cite{eomccsdbenchmark}. A quantitative agreement with the experiment, with a MAE of 0.07 eV, can be achieved with CVS-EOM-CCSDTQ\cite{eomccsdbenchmark}. Sub-electronvolt accuracy can also be achieved at a lower cost with second and third-order algebraic diagrammatic construction (ADC) methods within an intermediate state representation approach.\cite{schirmer_1982,schirmer_trofimov_2004,wenzel_dreuw_2016,adcstudy}

Large errors in response theory methods are usually attributed to the insufficient relaxation of molecular orbitals in the presence of a core hole. The second family of delta methods aims to properly account for orbital relaxation by explicitly optimizing the core-ionized state. In such calculations, the difference between the energy of the core-ionized and ground state is reported as the CEBE. Remarkably, even at the Hartree-Fock ($\Delta$HF) level of theory, the CEBEs can be calculated within 1 eV, an error comparable to the accuracy of CVS-EOM-CCSD.\cite{scfbenchmark} inclusion of correlation effects at the level of second-order Møller–Plesset perturbation theory, known as $\Delta$MP2 energies, can bring the mean absolute error to 0.5 eV.\cite{grabowski_2018} Further improvements, which bring the MAE below 0.2 eV, include the use of a specially calibrated basis set;\cite{mp2benchmark} application of the spin-component-scaled technique;\cite{grabowski_2018} or employment of restricted open-shell MP2 theory.\cite{hzyemp,neuscamman_2020} Recently developed square gradient minimization algorithm\cite{hait_sgm_2020} is capable of achieving similar precision by solving restricted open-shell Kohn-Sham equations.\cite{hait_2020}

So far, the highest accuracy method for computing CEBEs is the coupled-cluster theory,\cite{zheng_cheng_2019,mgh_2022} often referred to as the golden standard of computational chemistry. Until recently, however, the application of the $\Delta$CC method has been sparse primarily because of convergence issues: the creation of a hole in a core orbital opens room for core-to-virtual and valence-to-core double transitions that have very similar energies with different signs (sometimes called a near-degeneracy issue), resulting in the divergence of coupled-cluster equations. To circumvent these issues, inspired by the CVS scheme from response theory, \citeauthor{zheng_cheng_2019} in \citeyear{zheng_cheng_2019} have shown that exclusion of such troublesome transitions can solve convergence issues without significantly affecting the accuracy of results.\cite{zheng_cheng_2019} Additional schemes solving the near-degeneracy issue have been developed and tested by \citeauthor{mgh_2022} in \citeyear{mgh_2022}, who benchmarked the accuracy of $\Delta$CC calculations extrapolated to the CBS limit by using energies in the aug-cc-pCVnZ (heavy)/aug-cc-pVDZ (hydrogen) with $n=T, Q$ basis set for 18 organic molecules. Notably, these extrapolations are only approximations because calculation of the true CBS limit requires using $\Delta$CC results in pentuple, and sometimes even hexuple zeta basis sets. Such $\Delta$CC calculations, however, may be prohibitively expensive even in quadruple-zeta basis sets for large molecules because of high computational cost of CCSD\cite{purvis_bartlett_1982} and CCSD(T)\cite{pople_head-gordon_1989}, which scale as $O(N^6)$ and $O(N^7)$ respectively, where $N$ is the number of electrons in the system. A protocol that avoids these high costs without sacrificing accuracy is thus highly desirable.

In this work, we study the composite wave function based schemes that allow one to quantitatively (within 0.02 eV) recover the accuracy of $\Delta$CC calculations by correcting large basis $\Delta$MP2 calculations with a $\Delta$MP2-$\Delta$CC difference in a small basis. Because MP2 calculations do not suffer from convergence issues and are computationally less expensive, such composite schemes make predictions of XPS spectra for large molecules more feasible. We begin by benchmarking the accuracy (measured as mean absolute error, MAE) of $\Delta$HF, $\Delta$MP2, $\Delta$CCSD, and $\Delta$CCSD(T) at calculating CEBEs in basis sets of different sizes. We continue by analyzing the dependence of the performance of different extrapolations to the CBS limit of $\Delta$CCSD and $\Delta$CCSD(T) calculations on the sizes of the basis sets included in the extrapolation and uncover element-specific trends. We then investigate the impact of the choice of the large and small basis for the MP2-based composite extrapolation scheme. We conclude with practical recommendations for highly accurate (within 0.15 eV) ab initio prediction of K-edge CEBEs.

\section{Theory and computational details}
\textit{K}-edge CEBEs were calculated for carbon (26 molecules, hereafter referred to as C-series), nitrogen (30 molecules, N-series), oxygen (25 molecules, O-series), and fluorine (13 molecules, F-series) for a total of 94 data points. For some molecules, CEBEs were calculated for the ionization of different atoms within the molecule. Molecules (see Table S1 for a list) were selected from the table of experimental \textit{K}-edge CEBEs compiled by \citeauthor{experimentaldatabase}\cite{experimentaldatabase} with an attempt to include molecules of varying sizes and in which the ionized orbitals are in different chemical environments. Whenever multiple values were reported for one compound, an arithmetic average was used.

Experimental geometries from Computational Chemistry Comparison and Benchmark DataBase\cite{cccbdb} were used whenever available. If experimental geometries were unavailable, the results of full MP2 geometry optimizations with aug-cc-pVQZ or aug-cc-pVTZ basis sets were taken from the same database. If such results were not available, an RI-MP2\cite{orca_ri_mp2} geometry optimization was performed in cc-pVQZ\cite{dunning_HBNe,dunning-AlAr,orca-ccPVQZ/C} basis set (Table S1 specifies which geometries were used for each molecule). The geometry of a neutral species was used to calculate both ground and core-ionized state energy.

The calculations of core-ionization energies were performed in Dunning basis sets cc-pVnZ/cc-pCVnZ, with $n=D,T,Q,5$. In this mixed basis set, elements for which CEBEs are calculated are all treated in core-enhanced basis sets cc-pCVnZ,\cite{dunning-HBNe-core,dunning-AlAr-core} while the other elements were treated in cc-pVnZ.\cite{dunning_HBNe,dunning-AlAr} For example, in the calculation of a CEBE of one of the oxygen atoms in formic acid, all oxygen atoms were treated in cc-pCVnZ basis set, while carbon and hydrogen atoms were treated in a cc-pVnZ basis.

The core-electron binding energy (CEBE) in the delta-methods family is calculated as a difference between the single-point energy of the core-ionized and ground state (eq. \ref{eq:deltacalc}). 
\begin{equation}
 \label{eq:deltacalc}
 E_{\textsc{CEBE}} = E_{\textsc{ion}}-E_{\textsc{ground}}
\end{equation}
Energies and wave functions of the ground- and ionized states are calculated using Hartree-Fock (HF). The algorithm for the optimization of the ionized state is modified to proceed through the maximum overlap method (MOM)\cite{mom_paper} to avoid the variational collapse to the ground state. SCF equations for the ionized state are converged with the direct inversion in the iterative subspace\cite{diispaper} (see Table S2 for specification of DIIS parameters). Resulting HF wave functions are used as a starting point for calculating correlation energies with MP2, CCSD, and CCSD(T). The empty core orbital was excluded (implemented as freezing in PySCF\cite{pyscf_1,pyscf_2}) from amplitude calculations within MP2 and CC to improve their convergence. Finally, a 1-electron spin-free X2C approximation\cite{dyall_2001,liu_peng_2009} has been applied to all calculations to account for scalar relativistic effects.

If the molecule of interest contains symmetrically equivalent atoms, the core orbitals in the RHF solution are delocalized over those equivalent atoms, and a vacation of such orbital leads to the delocalization of the core hole as well. Practically, this results in an inaccurate orbital relaxation and an overestimation of core-ionization energies by more than 10 eV. The problem is resolved by applying a localization scheme to all atoms, e.g., the Boys localization,\cite{boys_1960} or by explicitly localizing the core orbitals of symmetrically identical atoms. 

As mentioned previously, precise extrapolation to the CBS limit of coupled-cluster energies is impractical due to the high computational cost of coupled-cluster calculations in large basis sets. Hence, lower-cost approximation schemes must be applied. One commonly used scheme is the two-point extrapolation from $\Delta$CC energies calculated in triple and quadruple-zeta basis sets using a two-parameter equation $E = a + bn^{-3}$ described by Helgaker.\cite{helgaker-1997} Energies from this scheme are denoted as X-Y-CCSD and X-Y-CCSD(T), in which $X$ and $Y$ refer to the size of the basis set based on which the extrapolation was made.

An alternative approach to approximate CBS values is to use a composite method (eq. \ref{eq:mp2gen}-\ref{eq:mp2diff}), which corrects MP2 energies in a large basis set with a difference between CC and MP2 energies in a small basis set.\cite{mp2scheme-2011} This technique utilizes the observed linear relationship between MP2 and CC energies, offset with a constant factor $\delta_{\text{CC-MP2}}$. This scheme has proven to be a good approximation for energies of non-covalent interactions in large systems.\cite{sherrill_2006,hobza_2009}
\begin{align}
 \label{eq:mp2gen}
 E_{\text{CC}}^{\text{large basis}} &\approx E_{\text{MP2}}^{\text{large basis}} + \delta_{\text{CC-MP2}}^{\text{small basis}} \\
 \delta_{\text{CC-MP2}}^{\text{small basis}} &= E_{\text{CC}}^{\text{small basis}} - E_{\text{MP2}}^{\text{small basis}} \label{eq:mp2diff}
\end{align}
We will refer to this scheme as MP2[X Y]+DifZ where \textit{X}, \textit{Y} denote the size of the basis set based on which the \textit{large basis} MP2 value was obtained, \textit{Z} refers to the size of the \textit{small basis} set, and DifZ or DifZ(T) denotes whether the difference in the \textit{small basis} was taken between CCSD and MP2 or CCSD(T) and MP2 respectively. Calculations of CEBEs were performed in cc-pCVnZ/cc-pVnZ (described above, $n = D, T, Q, 5$), STO-3G,\cite{hehre1969a,hehre1970a} STO-6G,\cite{hehre1969a,hehre1970a} 3,-21G,\cite{binkley1980a,gordon1982a} 4-31G,\cite{ditchfield1971a,hehre1972b,hehre1972c} and 6-31G.\cite{gordon1982a,ditchfield1971a,dill1975a,francl1982a,hehre1972a}

The latest (2.1.1) version of PySCF\cite{pyscf_1,pyscf_2} was used for all CEBE calculations. Geometry optimizations were performed in the latest (5.0.3) version of ORCA.\cite{orca_1,orca_2,orca_libint2,orca_libxc}

\section{Results and Discussion}
\subsection{Convergence of basis for HF, MP2, and CC CEBEs.}
\begin{figure}[h]
  \centering
  \includegraphics[width=\linewidth]{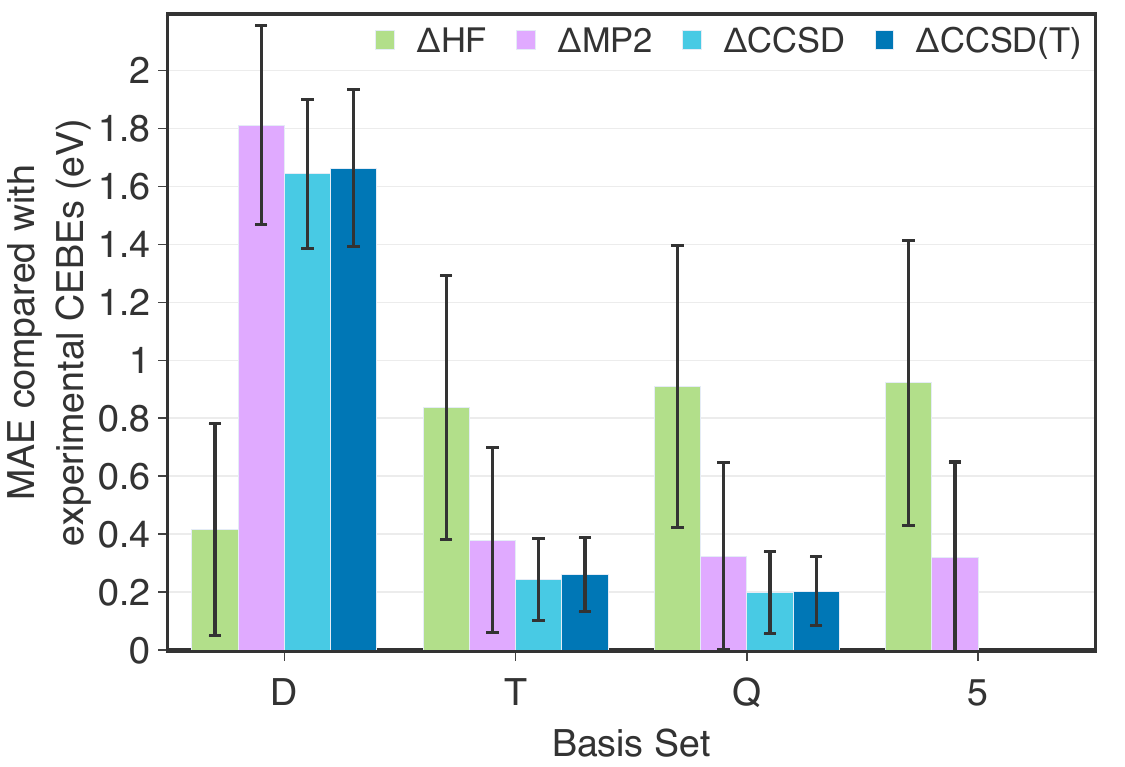}
  \caption{Accuracy (mean absolute error in eV relative to experimental values) of $\Delta$HF and $\Delta$post-HF methods in cc-pVXZ/cc-pCVXZ basis sets (X=D, T, Q, 5) at the calculation of 94 \textit{K}-edge CEBEs. Error bars show standard deviations of absolute errors.}
  \label{fig:methods_bars_all}
 \end{figure}

The mean absolute errors (MAE) for the CEBEs rely heavily on the basis set's size and the method. Remarkably, $\Delta$HF calculations in a double-zeta basis set produce CEBEs with a MAE of 0.41 eV, significantly lower than the MAE of $\Delta$MP2 (1.81 eV) or $\Delta$CCSD (1.64 eV) in the same basis (see Fig. \ref{fig:methods_bars_all}). Such a small error could be explained by fortunate cancellation of systematic errors during the calculation of energy difference between the ionized and ground states. Surprisingly, the MAE for $\Delta$HF increases with the basis set size: for triple, quadruple, and pentuple zeta, it is 0.84, 0.91, and 0.92 eV, respectively. In contrast, $\Delta$MP2 and $\Delta$CC calculations improve as the basis size increases (Table S3). As expected, $\Delta$CC calculations provide the most accurate CEBEs with a MAE of 0.20 eV in a quadruple basis set. Inclusion of a perturbative triples correction to $\Delta$CCSD result increases the MAE, a result in agreement with the work of \citeauthor{mgh_2022}.\cite{mgh_2022}

\begin{figure}[h]
  \centering
  \includegraphics[width=\linewidth]{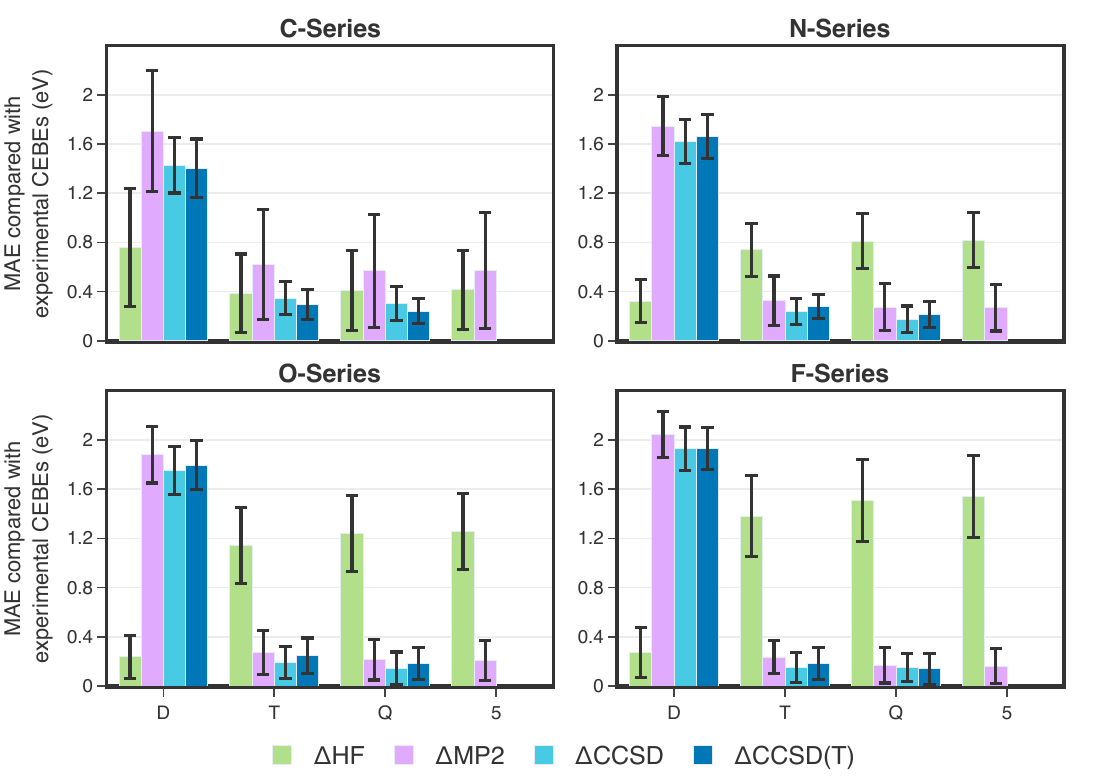}
  \caption{Accuracy (mean absolute error in eV relative to experimental values) of $\Delta$HF and $\Delta$post-HF methods in cc-pVXZ/cc-pCVXZ basis sets (X=D, T, Q, 5) at the calculation of core (\textit{K}-edge) ionization energies of electrons localized on carbon (25 molecules), nitrogen (31 molecules), oxygen (25 molecules), and fluorine (13 molecules). Error bars show standard deviations of absolute errors.}
  \label{fig:methods_bars_series}
 \end{figure}

When a similar analysis is performed for CEBEs grouped based on the element on which the ionized orbitals are located, it is observed that the trends differ between element series (Fig. \ref{fig:methods_bars_series} and Tables S4-S7). For example, the MAE for $\Delta$HF increases with the size of the basis set for N-, O-, and F-series but decreases for C-series. Another peculiarity of the carbon series is that $\Delta$HF CEBEs have smaller MAE than $\Delta$MP2 CEBEs in triple, quadruple, and pentuple bases. Finally, $\Delta$CCSD(T) energies (MAE 0.24 eV) are 20\% more accurate than $\Delta$CCSD energies (MAE 0.30 eV) for C-series, but 20-30\% less accurate for N- and O-series. The effect of perturbative triples for the F-series is ambiguous and dependent on the basis size. In effect, the lack of improvement upon introduction of perturbative triples, described by \citeauthor{mgh_2022},\cite{mgh_2022} seems to be an artifact of the selection of molecules for benchmark studies, which tend to have an abundance of oxygen-based ionizations.

\subsection{Choice of the basis sets for $\Delta$CC extrapolations.}
\begin{figure}[h]
   \centering
   \includegraphics[width=\linewidth]{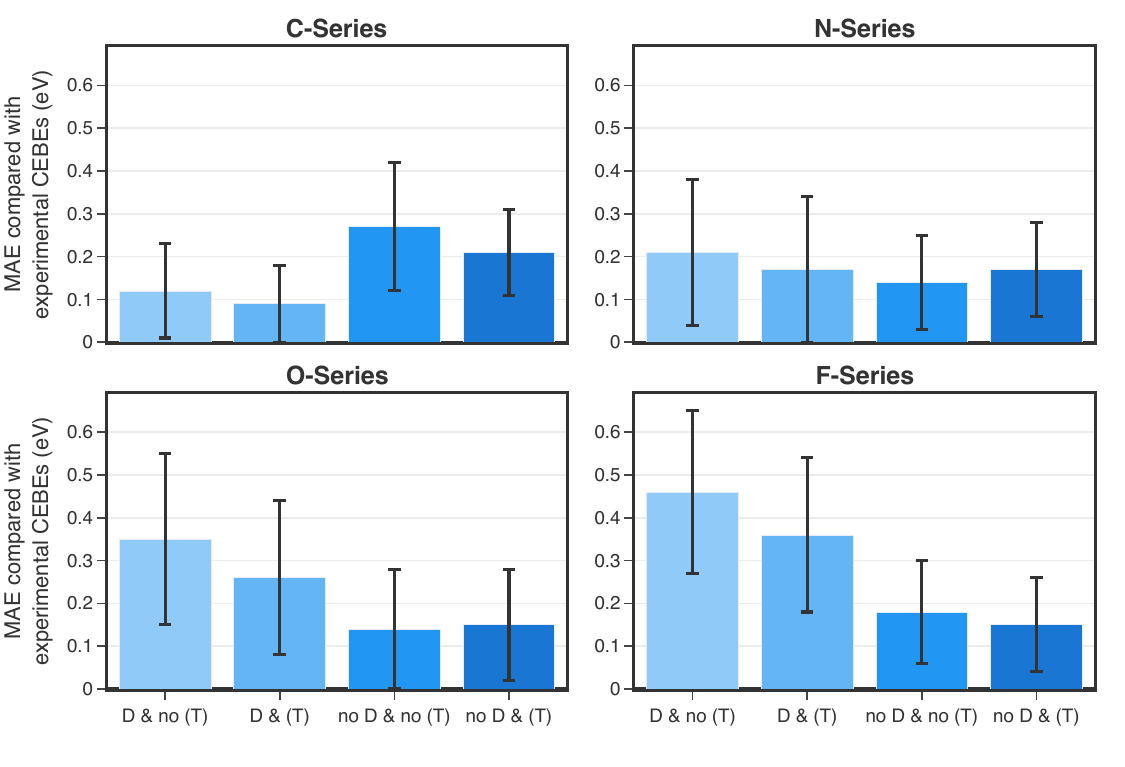}
   \caption{Mean absolute errors (in eV) of D-T-Q-CCSD, D-T-Q-CCSD(T), T-Q-CCSD, and T-Q-CCSD(T) extrapolations at the calculation of core (\textit{K}-edge) ionization energies of electrons localized on carbon (25 molecules), nitrogen (31 molecules), oxygen (25 molecules), and fluorine (13 molecules). Error bars show standard deviations of absolute errors.}
   \label{fig:bars_dtstudy-ccsd}
  \end{figure}

Averaged over all molecules, two-point extrapolations T-Q-CCSD and T-Q-CCSD(T) result in a MAE of 0.18 and 0.17 eV, respectively (Table S8). The inclusion of double-zeta results worsens the accuracy: D-T-Q-CCSD and D-T-Q-CCSD(T) have a MAE of 0.26 and 0.20 eV, respectively. However, as seen in Fig. \ref{fig:bars_dtstudy-ccsd} and Tables S9-S12, these \textit{global} averages hide the element-specific trends. For example, the inclusion of double zeta results for carbon series reduces the MAE by a factor of two from 0.27 to 0.12 eV for T-Q-CCSD vs. D-T-Q-CCSD and from 0.21 to 0.09 eV for T-Q-CCSD(T) vs. D-T-Q-CCSD(T). In contrast, for nitrogen series, D-T-Q-CCSD is 50\% less accurate than T-Q-CCSD (0.21 vs 0.14 eV respectively). The systematic nature of the increase in accuracy upon inclusion (for C-series) or exclusion (for all other series) of double zeta results is confirmed by correlation plots (Fig. S5-6). It should be noted that such element-specific impact of double zeta basis may be an artifact of the differences in the parametrization of double zeta basis for different elements in the Dunning family. Nonetheless, because Dunning basis sets are a predominant (if not exclusive) choice for $\Delta$CC calculations, such effect bears practical importance. For all series, the effect of triples correction is marginal and is more significant when suboptimal (for example, including double zeta results when it is better not to) extrapolation schemes are used. Notably, the effect of triples is systematic, here defined as having the same sign and similar in relative magnitude for all molecules in a series, for molecules within N-, O-, and F-series (Fig. S9-10), but is irregular for carbon series: while CEBEs for all but 3 molecules is improved upon inclusion of perturbative triples, the magnitude of the change is not uniform.

\subsection{MP2-based composite extrapolation scheme}
Extrapolation of $\Delta$MP2 CEBEs to the CBS limit does not result in a significant increase of accuracy: the MAE for $\Delta$MP2 values extrapolated from results in $2,3,4,5$-zeta basis sets is 0.26 eV (Tables S13-S16). These results are improved significantly for carbon and nitrogen series if a $\delta_{\text{CC-MP2}}^{\text{D}}$ correction is introduced in the double-zeta polarized basis (D refers to cc-pCVDZ/cc-pVDZ basis): the MAE lowers to 0.17 eV (Table S18). Greater accuracy can be achieved if extrapolation schemes are chosen differently for each series: the errors can be then reduced to 0.10 eV for C-series (Table S19), 0.14 eV for N-series (Table S20), 0.13 eV for O-series (Table S21), and 0.15 eV for F-series (Table S22). 
\begin{figure}[h]
  \centering
  \includegraphics[width=\linewidth]{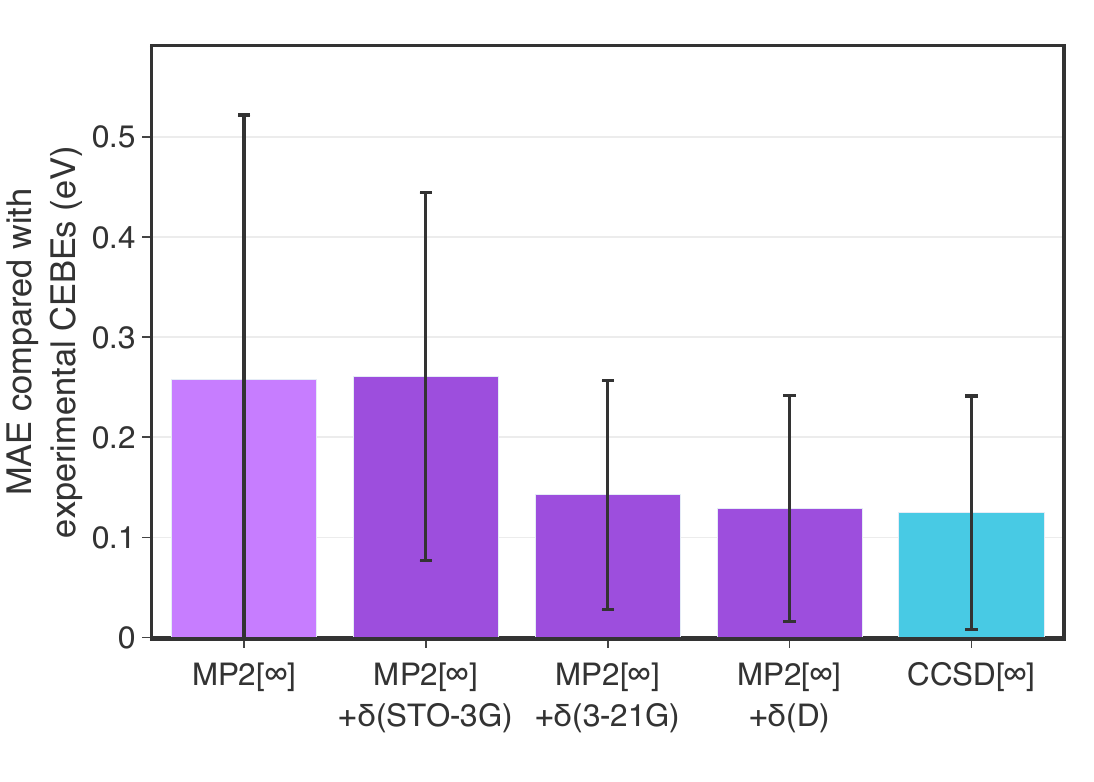}
  \caption{Mean absolute errors (in eV) of MP2-based extrapolation schemes in comparison to CCSD-based extrapolation. Calculations in 2-$\zeta$ are included only for C-based molecules. MP2 extrapolations do not include results in pentuple basis. Perturbative triples are included for C- and F-based molecules. Error bars show standard deviations of absolute errors.}
  \label{fig:bars_summary-no5}
 \end{figure}

 \begin{figure}[h]
  \centering
  \includegraphics[width=\linewidth]{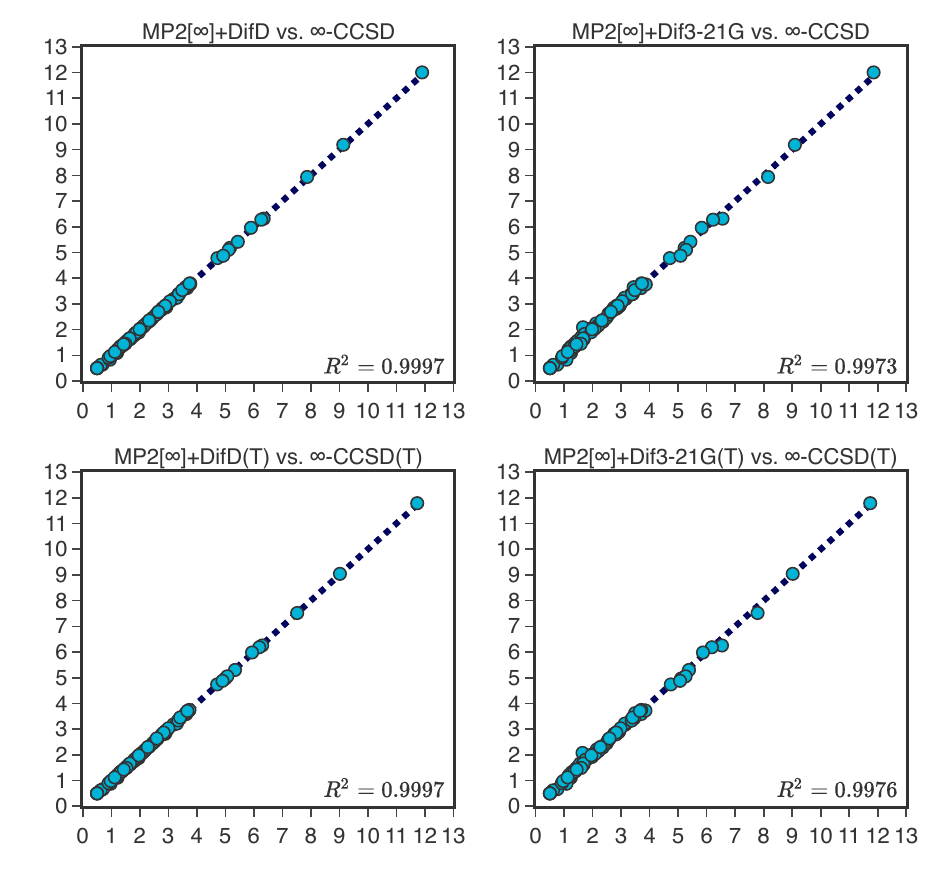}
  \caption{Comparisons of CEBEs for all 94 molecules in the study calculated with different extrapolations to the CBS limit. Energies are reported as shifts (in eV) from the lowest value in the element-specific series (set to equal 0.5 eV). The values from MP2-based extrapolation are reported on the x-axis, and CCSD-based extrapolation is reported on the y-axis. The dotted line is the curve $y=x$. The 2-$\zeta$ calculations are included only for carbon-based molecules. MP2 extrapolations do not include results in pentuple basis.}
  \label{fig:correlate_summary}
 \end{figure}

In the previous subsection, we have demonstrated (based on $\Delta$CC CEBEs) that it is optimal to include double zeta results only in extrapolations for carbon-based molecules and include perturbative triples correction in carbon- and fluorine-based molecules. Following these insights, hereafter, we will use MP2[D T Q]+DifZ(T) scheme for C-series, MP2[T Q]+DifZ for N- and O-series, and MP2[T Q]+DifZ(T) for F-series (the MP2-based versions of Fig. 3 are Fig. S2-3). As seen in Fig. \ref{fig:bars_summary-no5}, such an approach results in a MAE for an MP2-based composite extrapolation scheme of 0.128 eV, practically identical to the MAE from CC-based extrapolations of 0.125 eV, when the $\delta_{\text{CC-MP2}}$ correction is calculated in cc-pCVDZ/cc-pVDZ. Notably, if a significantly smaller regular double zeta basis (3-21G) is used to calculate the $\delta_{\text{CC-MP2}}$ correction, only a marginally larger MAE of 0.142 eV is observed. 

The CC- and MP2-based extrapolation schemes are similar not only on average but on a per-molecule basis, as seen in Fig. \ref{fig:correlate_summary}. In fact, MP2[$\infty$]+DifD recovers $\Delta$CC-based CBS values within 0.04 eV (as measured by RMSE) and 0.03 eV (as measured by MAE), while MP2[$\infty$]+DifD(T) recovers $\Delta$CC CBS values within 0.03 eV (RMSE) \& 0.02 (MAE). Such small differences establish a quantitative equivalency between the MP2- and CC-based extrapolation schemes. If the correction is calculated in a regular double-zeta basis, i.e., 3-21G, an MP2-based scheme recovers CC-based values within 0.10 eV (RMSE) \& 0.07 eV (MAE) and 0.09 eV (RMSE) \& 0.06 eV (MAE) without and with perturbative triples respectively. These results are remarkable as $\Delta$MP2 calculations are both asymptotically and practically faster given that MP2 scales as $O(N^5)$ and it only requires a single computation of perturbative correction, unlike the inherently iterative coupled-cluster calculations.

\subsubsection{Effect of the size of the large basis on the accuracy of composite schemes}
\begin{figure}[h]
  \centering
  \includegraphics[width=\linewidth]{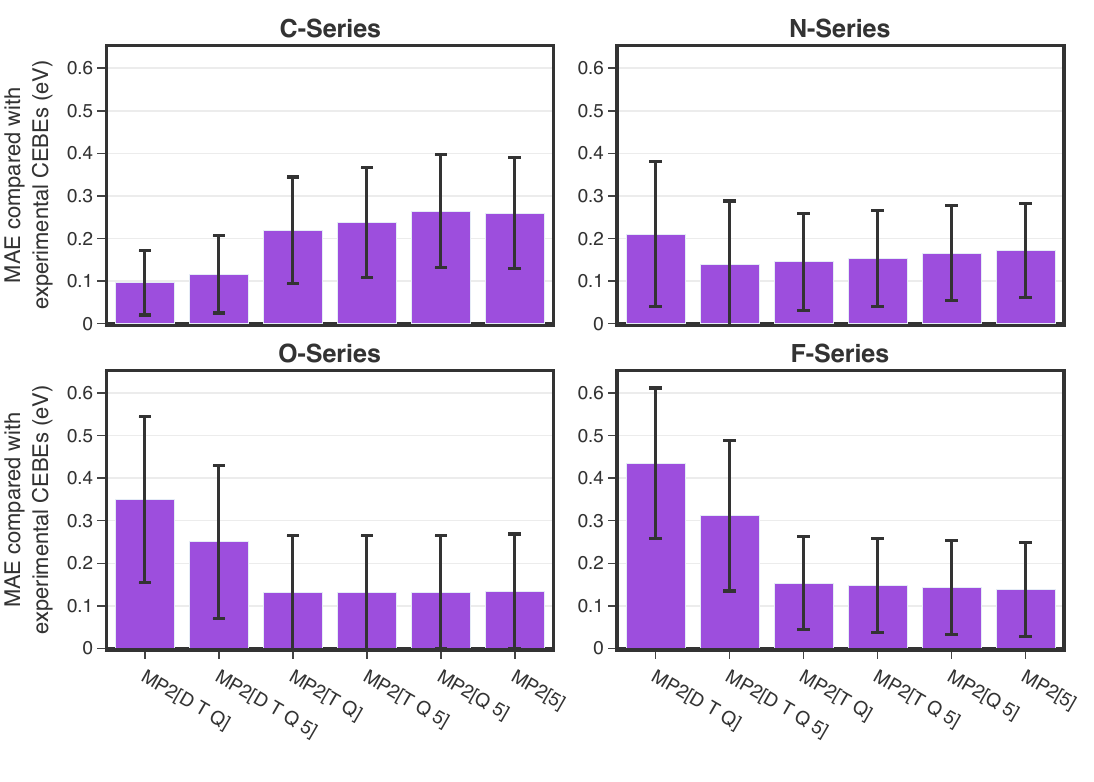}
  \caption{Mean absolute errors (in eV) of MP2-based extrapolations using different extrapolations of $E_{\text{MP2}}^{\text{large basis}}$ value for the calculation of core (\textit{K}-edge) ionization energies of electrons localized on carbon (25 molecules), nitrogen (31 molecules), oxygen (25 molecules), and fluorine (13 molecules). The $\delta_{\text{CC-MP2}}^{\text{D}}$ correction is calculated with cc-pVDZ/cc-pCVDZ basis set. Error bars show standard deviations of absolute errors.}
  \label{fig:mp2_big}
 \end{figure}

Equations \ref{eq:mp2gen}-\ref{eq:mp2diff} require the calculation of a $\Delta$MP2 CEBE in a large basis set. To establish the quantitative equivalence of the MP2-based extrapolation scheme, we have used MP2 results in double (for C-series only), triple, and quadruple zeta basis sets. Two questions warrant further investigation: 1. Can the accuracy be improved even further if MP2 results in the pentuple basis set are included? 2. Can we use fewer basis sets in MP2 extrapolation? Somewhat surprisingly, as seen in Fig. \ref{fig:mp2_big} and Tables S19-S22, the results in pentuple zeta basis either do not lower the MAE at all (as for oxygen and fluorine series) or even slightly increase it. Perhaps even more surprisingly, the MP2[T Q]+DifD, MP2[T Q 5]+DifD, MP2[Q 5]+DifD, and MP2[5]+DifD all result in practically identical MAEs for oxygen and fluorine series. The MAEs for nitrogen series increase slightly as the number of basis sets included in MP2 extrapolation decreases: from 0.14 eV for MP2[T Q]+DifD to 0.18 eV for MP2[Q]+DifD. In other words, accurate CEBE predictions can be made with less computationally expensive calculations if extrapolation schemes are chosen separately for each ionized element.

\subsubsection{Effect of the size of the small basis on the accuracy of composite schemes}
\begin{figure}[h]
  \centering
  \includegraphics[width=\linewidth]{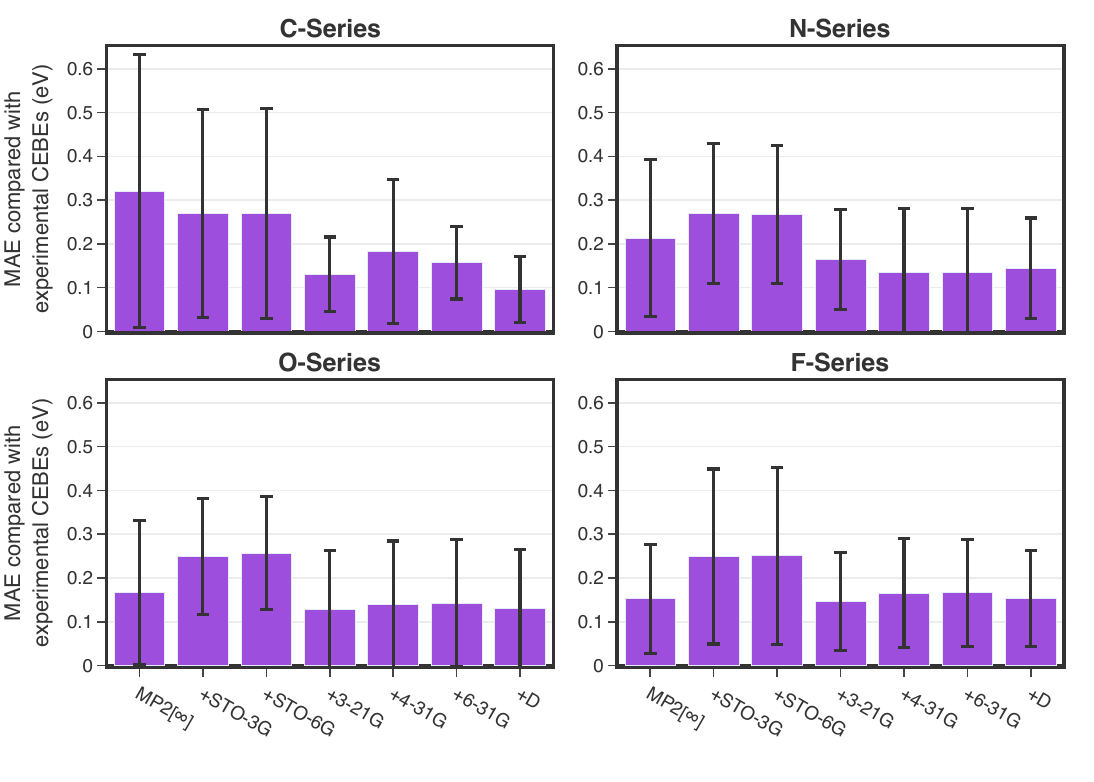}
  \caption{Mean absolute errors (in eV) of MP2-based extrapolations using different $\delta_{\text{CC-MP2}}$ corrections for the calculation of (\textit{K}-edge) ionization energies of electrons localized on carbon (25 molecules), nitrogen (31 molecules), oxygen (25 molecules), and fluorine (13 molecules). The $E_{\text{MP2}}^{\text{large}}$ value is calculated with $2,3,4$-$\zeta$ for C-series and $3,4$-$\zeta$ for C-, O-, and F-series. Error bars show standard deviations of absolute errors.}
  \label{fig:mp2_small}
 \end{figure}

We now investigate the question of how small should be the basis set in the $\delta_{\text{CC-MP2}}^{\text{small basis}}$ correction in the equation \ref{eq:mp2gen}. As seen in Fig. \ref{fig:mp2_small}, the minimal atomic orbital basis sets (such as STO-3G or STO-6G) are insufficiently large for the use in MP2-based extrapolation schemes as the MAEs are roughly twice as large as when cc-pCVDZ/cc-pVDZ is used. In contrast, 3-21G, 4-31G and 6-31G all result in comparable accuracy.

Given the results in Fig. \ref{fig:mp2_small} and Tables S18-S22, \textit{K}-edge CEBEs can be predicted ab initio within 0.19 eV by simply using the MP2[Q]+DifD scheme, involving an MP2 calculation in a double zeta and quadruple zeta basis sets and a single CCSD calculation in the double zeta basis. These results are effectively identical to the significantly more expensive T-Q-CCSD calculations. To improve the accuracy of ab initio predictions, one must employ element-specific extrapolation schemes: MP2[D T Q]+DifD scheme for C-series and MP2[T Q]+DifD for N-, O-, and F-series. If one is willing to sacrifice up to 10\% accuracy in favor of a lower computational cost, the small basis correction can be calculated in 3-21G instead.

\section{Conclusions}
This paper reports a systematic element-specific study of the impact of basis set selection for extrapolations to the CBS limit both for $\Delta$MP2 and $\Delta$CC calculations. Firstly, we reproduce the conclusions of \citeauthor{zheng_cheng_2019}\cite{zheng_cheng_2019} and \citeauthor{mgh_2022}\cite{mgh_2022} that T-Q-CCSD is the best extrapolation scheme on average, but we note that this result hides element-specific trends: we found that carbon-based CEBEs are predicted significantly more accurately if energies from double zeta basis are included in the extrapolation. This result is important both for experimentalists, especially those who are mostly concerned with carbon-based XPS, and theoreticians, who select molecules for benchmark studies, as the performance of a given method on average could depend not only on the quality of the method itself but also on the relative proportions of elements that are ionized in the dataset. We also find that $\Delta$MP2 extrapolated to the CBS limit corrected by $\delta_{\text{CC-MP2}}^{\text{small basis}}$ quantitatively reproduces optimally extrapolated $\Delta$CC CEBEs at a fraction of their computational cost. For example, MP2[T Q]+DifD can be used to estimate CEBEs of molecules for which T-Q-CCSD calculations are prohibitively expensive. Just as with $\Delta$CC extrapolations, the choice of the basis for the MP2-based composite method is element-specific, so the inclusion of a double-zeta basis in the extrapolation of $\Delta$MP2 results will be useful for carbon-based ionizations. Finally, this work shows that neither perturbative triples nor MP2 calculations in pentuple basis set systematically and significantly improve the quality of predictions. Our results suggest that highly accurate ab initio prediction of the XPS spectra of large molecules is feasible with currently available methods.

Future work may be done in three directions. First, a more efficient implementation of MP2 or CCSD can be used to quantitatively assess the practical computational costs of proposed composite MP2-based extrapolation schemes relative to CCSD calculations in triple and quadruple basis sets. An investigation of the change in accuracy if MP2 results are replaced with a resolution of identity (RI) approximation of MP2 (RI-MP2) can also be performed. It would also be beneficial to search for ways to eliminate dependence on iterative $\Delta$CC results altogether. Second, an investigation of the accuracy of MP2-based extrapolation schemes for third-row elements and transition metals is of great practical importance. Third and finally, all calculations reported in this paper relied upon manual freezing of core orbitals to avoid the near-degeneracy issue of coupled-cluster calculations. A black box implementation is especially desirable for L-edge CEBEs and beyond as the density of orbitals increases with the principal quantum number.

\begin{acknowledgement}
We thank National Science Foundation for funding this project (CHE-2154938). Anton Morgunov and Yu-Che Chien are grateful for the additional financial support by the Undergraduate Research Opportunities Program (UROP) at Massachusetts Institute of Technology.
\end{acknowledgement}

\begin{suppinfo}
The following files are available free of charge.
\begin{itemize}
  \item Supporting Information: addditional tables and figures for the statistical metrics for different methods and extrapolation schemes
  \item \url{https://github.com/anmorgunov/cebe_prediction}: code used to run calculations, analyze the results, and generate the tables and figures in this paper. Also includes the CEBE values for all molecules in the study.
\end{itemize}

\end{suppinfo}

\bibliography{cebe_preprint_v4}

\end{document}